\documentclass[letterpaper,twocolumn,10pt]{article}
\usepackage{usenix,epsfig,endnotes}
\usepackage{multirow}
\usepackage{booktabs}

\begin{document}

%don't want date printed
\date{}

%make title bold and 14 pt font (Latex default is non-bold, 16 pt)
\title{\vspace*{-1cm}\Large \bf Old Wine in New Skins?  Revisiting the Software Architecture 
 \newline for IP Network Stacks on Constrained IoT Devices}

%for single author (just remove % characters)
\author{
{\rm Hauke~Petersen}, {\rm Martine~Lenders}, {\rm Matthias~W\"ahlisch}\\
Freie Universit\"at Berlin, Germany
\and
{\rm Oliver~Hahm}, {\rm Emmanuel~Baccelli}\\
INRIA, France
% copy the following lines to add more authors
% \and
% {\rm Name}\\
%Name Institution
} % end author

\maketitle

% Use the following at camera-ready time to suppress page numbers.
% Comment it out when you first submit the paper for review.
%\thispagestyle{empty}

\subsection*{Abstract}
In this paper, we argue that existing concepts for the design and implementation of network stacks for constrained devices do not comply with the requirements of current and upcoming Internet of Things (IoT) use cases. The IoT requires not only a lightweight but also a modular network stack, based on standards. We discuss functional and non-functional requirements for the software architecture of the network stack on constrained IoT devices. Then, revisiting concepts from the early Internet as well as current implementations, we propose a future-proof alternative to existing IoT network stack architectures, and provide an initial evaluation of this proposal based on its implementation running on top of state-of-the-art IoT operating system and hardware.

%\begin{IEEEkeywords}
%IEEEtran, journal, \LaTeX, paper, template.
%\end{IEEEkeywords}

\section{Introduction}
\label{sec:introduction}

% the IoT vision
The Internet of Things (IoT) promises a future where all machines have started talking to one another, including billions of cheap, tiny, programmable, communicating devices (aka Things) such as wired or wireless sensors, and actuators. Based on various types of low-cost microcontrollers and communication chips, those devices will significantly increase heterogeneity within the Internet. 

The past has shown that the success of the Internet depends on the availability of network stack(s) that allow for flexible composition of standards and enabling a wide variety of optional features to fit heterogeneous use cases.

The design and implementation of networks stacks challenged system engineers since the very beginning of computer networking. At that time computers exhibited severe hardware resources constraints, similar to IoT devices nowadays in terms of memory (kBytes instead of GBytes), and in terms of CPU (Mflops instead of Gflops). However, in contrast to the 80s and early 90s, the heterogeneity and thus the set of options and scenarios is much larger in the IoT, which increases complexity~\cite{RFC-5867, RFC-5826, RFC-5673, RFC-5548}. Furthermore, Moore's Law does not apply to microcontrollers, and thus, such tiny devices will remain prevalent in the future~\cite{RFC-7228}. Since full-featured systems such as Linux cannot be accommodated on such tiny devices, novel solutions are needed.

In this paper, we argue to revisit the design and implementation space of network stacks for constrained devices. Recent operating systems (e.g., RIOT~\cite{hahm2014}) support Linux-like functions but comply with the hardware constraints of IoT hardware, which gives potential to build flexible network stacks with low memory foot-print.

We target class 1 devices \cite{RFC-7228} or bigger, i.e., devices with at least 10~kByte of RAM and a few tens of kByte of ROM. We believe that for even more constrained devices, there is no way around specialized, simplified, and highly optimized implementations. Therefore, note that our goal is not to engineer the most memory-efficient network stack but to design a clean, structured, and universal network stack that can be reused for many different IoT use cases, while still being able to cope with constrained environments. In detail, our contributions are as follows:

\begin{enumerate}
\setlength{\itemsep}{-0.15cm}
\item we identify functional and non-functional requirements for the software architecture of the network stack on IoT devices (see Section~\ref{sec:objectives}),
\item we analyze existing IoT network stack architectures, from a systems point of view (see Section~\ref{sec:relatedwork}),%, and discuss assumptions for the targeted software platform (see \S~\ref{sec:assumptions}),
\item we propose an alternative architecture which we argue is more future-proof than existing architectures, because easier to use for continuous extensions, to configure for different IoT use cases, leveraging cleaner interfaces and newly available IoT operating system services (see Section~\ref{sec:architecture}),
\item we provide initial evaluation of the proposed architecture and show it complies with typical resource constraints of IoT hardware (see Section~\ref{sec:evaluation}).
%\item we provide initial evaluation of the proposed architecture and show it complies with typical resource constraints of IoT hardware, while keeping the complexity comparable to Linux (see \S~\ref{sec:evaluation}).
\end{enumerate}

\section{Assumptions \& Design Objectives}
\label{sec:objectives}

%Before we describe the architecture of the proposed network stack in detail we will point out a number of assumptions on the targeted software platform. These are important as they do not only influence implementation details but also have an impact on architectural design decisions.

As the complexity of software for embedded devices has increased over the last decade, it has become state-of-the-art to use operating systems even on memory and CPU constrained machines, such as IoT devices. 
A full-featured network stack is one of the most complex pieces of software to run on an embedded platform. 
By full-featured, we refer to a stack allowing for a complete implementation of the specifications per design.
This point is especially important, as one can easily simplify parts of an implementation at the price of limiting the extent of completeness that this implementation can achieve in the end. In the following, we will make the assumptions listed below.

%A full-featured network stack is one of the most complex pieces of software to run on an embedded platform. By full-featured, we mean that (i) the stack should comply with protocol standards specifications, and (ii) by design, it should allow for a complete implementation of the specifications. The latter point is especially important, as one can easily simplify parts of an implementation at the price of limiting the extent2" of completeness that this implementation can achieve in the end.

%Modern embedded operating systems (OS) do not only provide simple run-time environments in form of (real-time) task scheduling and synchronization.They also offer unified hardware abstraction for heterogeneous micro-controller platforms, driver models, software libraries, and testing environments.These features speed up the development and testing of applications on IoT devices and improve application portability. 
%not only unified hardware abstraction for heterogeneous microcontroller platforms but also offer (real-time) task switching and synchronization. These features speed up software development and enable a platform-independent design while still being able to cope with resource constraints. 

\textbf{Multi-Process \& Hardware Independence.}\quad
We assume that the network stack is built atop such an OS that provides the following features: (i) support of threads/processes, (ii) a lightweight process model, (iii) efficient inter-process communication (IPC), (iv) lightweight hardware abstraction, (v) a clean driver model, and (vi) a memory foot-print suitable for IoT devices. Assumptions (i)-(iii) allow for a modular network stack that is split over multiple processes without a significant overhead through administrative data structures and run-time drawbacks. Assumptions (iv) and (v) enable the network stack to be independent from specific IoT hardware platforms and network devices. We also assume that the OS allows the network stack to be open source, maintained by a lively community (similarly to Linux).

%\textbf{Open-Source \& Standard Programming.}\quad
%The second set of assumptions we make is that the OS allows the network stack to be (i) open source, with a lively community, (ii) implemented using standard programming paradigms and languages, (iii) built with standard tools (e.g., gcc), and (iv) debugged with standard tools (e.g., gdb). Resorting to exotic languages and tools is a dangerous gamble in the long run, because chances are high that they will remain exotic, and prevent the network stack to be developed and maintained by a critical mass of developers, repelled by the steep learning curve. A typical example of this phenomenon is TinyOS with the nesC language \cite{levis2012}.
It is worth noting that our assumptions are reasonable; the operating system RIOT~\cite{hahm2014} matches all of them and thus allows us a proof of concept of our proposed architecture.

\iffalse
\begin{itemize}
\item support of threads/processes
\item a light-weight process model
\item efficient inter-process communication (IPC)
\item light-weight hardware abstraction
\item a clean driver model
\end{itemize}
\fi

\subsection{High-Level Objective \& Approach}

The usual approach to deal with the heterogeneity of embedded systems in the IoT (i.e. hardware constraints, use cases) is to implement multiple network stacks --- each designed for a specific setup.
While this yields optimized results for a small group of scenarios, there are drawbacks: multiple implementations vastly increase efforts for implementation, testing, maintenance, and incur extra efforts to ensure interoperability.

%There are different deployment scenarios for a network stack in the IoT. From a networking perspective they range from simple sensor nodes to border routers, from simple leaf nodes in a routing tree to nodes utilizing multiple network interfaces. The same is true from a resource perspective: IoT devices range from class 0 devices~\cite{rfc7228} with a single digit number of kilobytes of RAM and ROM to class 1 and 2 devices with tens of kilobytes of RAM and ROM, or more.

%Commonly, to deal with this heterogeneity, multiple network stacks are specifically designed and implemented. This approach may help when focusing on a single use case but implies drawbacks in general: multiple implementations of network stacks increase efforts for implementing, testing, and maintaining each code-base, as well as more specific tests and effort to guarantee interoperability between stacks.

We thus pursue a different approach: we aim for a single, full-featured network stack that is flexible enough to work in a broad range of IoT scenarios, while still being efficient and small enough to run on constrained and battery-driven devices. In the following, we break down the various aspects of this high-level objective.
%categories of functional and non-functional requirements, each of which is discussed in relation with the conventional design space.

\subsection{Functional Requirements}

\textbf{\indent Focus on IPv6.}\quad
The network stack should enable end-to-end connectivity between IoT devices and any other Internet device. IPv6 is a good candidate for this functionality, together with the 6LoWPAN suite of IP protocols for low-power lossy networks (including RPL, UDP, CoAP etc.). Note that that our design should also easily apply to other layered network stacks. For this reason we will focus, but not exclusively, on IPv6.

%A design goal of this network stack is to enable constrained devices to connect directly to the internet. In contrary to gateway based silos, this means that each device can technically enter into end-to-end connectivity to any other device in the internet. This direct connection implies the use of routable internet addresses. When talking about billions of devices in the IoT it is obvious that the IPv4 address space is not able to cope with this number of devices. For this reason we will focus, but not exclusively, on IPv6 technology. The core of the network stack shall be formed by an IPv6 network layer and IPv6's adaption for constrained networks, 6LoWPAN.

\textbf{Full-featured.}\quad
We aim for a full-featured network stack in a sense that supported protocols should implement their specifications completely as a long-term goal.
The point is to prohibit design decisions which will limit future extensions of an implementation.
The rationale behind this is to allow for a generic solution, which can be tailored to fit various use cases, instead of a solution that is too specific by design. 

%The network stack modules should not only be 100\% standards compliant, but should also implement the specifications completely. If completeness is not fully achieved, the architecture must by design provision for full completeness (i.e., not prevent it). The rationale behind this is that the range of IoT use cases is wide, and maintaining multiple specialized network stacks should be avoided. Therefore a single, flexible, full-featured code base is preferable. 

\textbf{Support for multiple network interfaces.}\quad
IoT scenarios do not only include basic sensors with a microcontroller and a single low-power radio, but also border routers with multiple interfaces (e.g., Ethernet and IEEE 802.15.4) as well as upcoming IoT devices, which are likely to have multiple radio interfaces (e.g., IEEE 802.15.4 and Bluetooth). Thus, the network stack must be able to handle multiple network interfaces, and we argue that, if designed carefully, the overhead of multi-interface support is negligible compared to single interface support, even on constrained devices.

 %Nodes in classical sensor networks consist typically of some kind of resource constraint micro-controller with a single low-power radio. When looking at upcoming IoT devices we see a trend towards nodes with multiple interfaces. On the high-end side there are modern smart phones with typical 3 or more radio interface (i.e. 3G, IEEE802.11, Bluetooth, NFC). But also on the low-end side we see an increasing number of devices supporting more then one radio or network interface (i.e. 802.15.4, Bluetooth Smart, Ethernet). As the goal is further to use the same network stack for boarder routes as for leaf nodes the network stack needs to be able to handle multiple network interfaces. 

%One might object that support for multiple network interfaces is overkill for constrained devices and is not feasible with respect to the additional resources needed. We argue the contrary: by implementing the multi-interface support in an efficient way, we keep the overhead for each additional device to a minimum which leads to a well scaling solution. When configuring the network stack to only use a single interface the overhead is neglectable compared to single interface design. Adding support for multiple interfaces to a network stack later on is hard to do in a clean and maintainable way, configuring a multi-interface supporting network stack to use only a single interface can be done in an efficient way.

\textbf{Parallel data handling.}\quad
Most embedded network stacks achieve their small memory footprint by reducing their functionality, to the point where they are only able to handle a single network packet at a time. While this might be reasonable in some use cases, this is unrealistic in general. In particular, using IPv6 over spontaneous wireless networking, multiple services run in parallel, e.g.,  both routing and neighbor discovery protocols are tightly coupled to data transfers between nodes. Thus, the network stack must be able to handle multiple packets and data streams in parallel.

\subsection{Non-functional Requirements}

\textbf{\indent Open Standards and tools.}\quad
Decades of experience with the Internet indicate that deployment success depends on (open) standards.
To achieve future-proof interoperability despite heterogeneity amongst IoT devices, the network stack must be standard compliant. Heterogeneity is not only found in IoT hardware but also in development environments and processes. We argue that a standard network stack should only depend on open tools and standard paradigms (e.g. ANSI C) to allow easy integration.
Exotic tools and programming languages become a fatal hurdle on the way to reaching the critical mass of developers necessary to develop and maintain in the long run a piece of software as sophisticated as a network stack (a typical example of this phenomenon is TinyOS with the nesC language \cite{levis2012})

%Aside the variety in IoT hardware, heterogeneity can also be found in software environments used to develop code running on IoT devices. We argue that a standard IoT network stack should only depend on widely used open tools, and stick to well-known programming paradigms. Exotic tools and programming languages become a fatal hurdle on the way to reaching the critical mass of developers necessary to develop and maintain in the long run a piece of software as sophisticated as a network stack. 

\textbf{Configurability.}\quad
The objective is the design of a versatile network stack that can be adapted to a variety of IoT scenarios. 
However, the granularity of configuration should avoid too many configuration options that have unclear meaning and effects (and thus are only usable for experts). 
Key configuration parameters must be well documented and accessible from a central point to achieve a user-friendly and flexible solution.

\textbf{Extensibility via clean interfaces.}\quad
Clean interfaces yield two important advantages. First, it focuses modules on their core functionality, thus preventing entangled code. Second, it yields testability by design. Furthermore, modules and clean interfaces enable substitution of parts of the network stack, which can easily be tailored according to the IoT scenario. For example, it is straightforward to switch between two different implementations of a neighbor cache, one being optimized for run-time performance using a heap data structure, and another being optimized for memory efficiency using a simple circular list. However, again, the granularity of modules should remain coarse enough to avoid the pitfalls of ultra-fragmented code, which quickly becomes unmanageable, as analyzed in \cite{levis2012}.

\textbf{Low memory footprint.}\quad
%Memory (both flash memory and RAM) is the biggest cost driver for microcontrollers, thus memory footprint should be kept on a leash. 
While we do not aim for the smallest possible memory footprint (we have other goals, as stated above), we aim for very limited resources. 
For a concrete upper bound we aim for a maximum of 30Kb of ROM and 10Kb of RAM for a single interface configuration running 6LoWPAN, RPL and UDP.
These target numbers align well with the available resource on class 1 devices~\cite{RFC-7228}, which we expect to one of the most significant classes of IoT devices in the near future.

%For example, for a single interface configurations running 6LoWPAN and RPL, the target is to make the network stack use less than 10Kb of RAM and 30Kb of ROM. In particular, this target is aligned with available resources on class 1 devices~\cite{rfc7228}, which we expect to be the bulk of IoT devices in the near-future.

\textbf{Low-power design.}\quad
Many IoT devices are expected to run for years on small batteries. Experience shows that optimizations for low-power are harder to add on, and thus should built-in by design, from the very beginning. This has mainly two consequences: (i) the design of the network stack must allow to easily vary the protocols used in different scenarios, as best suited, and (ii) the implementation must use efficient data-structures and algorithms allowing maximum sleep intervals for the CPU.

\section{Related Work: Existing Network Stacks}
\label{sec:relatedwork}

Today's Internet is unthinkable without Linux/Unix and their network stacks, successors of the BSD 4.4 network stack~\cite{chesson1975, wehrle2004}.
Although they were originally developed in times when the memory constraints of a typical computer were roughly comparable with that of current IoT devices, their development followed fundamentally different design objectives, focusing predominantly on throughput (this manifests itself e.g., in the way buffers are designed).
Over the years this lead to a drastically increased memory footprint and made it inconceivable to run or port these stacks to IoT devices.

Over the last decade, and even more since 6LoWPAN has evolved, a number of network stacks have been developed specifically for embedded devices.
One category of stacks are ultra-minimalistic implementations, such as the work by \emph{Santos et al.} ~\cite{santos2013}, which -- by design -- are not extensible and cannot become a full-featured IP stack.
Thus, they do not meet the requirements from Section~\ref{sec:objectives}.
Various other stacks, as presented by several surveys, can be roughly be put in three groups (i) discontinued, (ii) proprietary and closed-source, or (iii) open-source and freely available~\cite{mazzer2009, sarwar2010, yibo2011}.
In the following we will focus on the third group (the analyzed requirements disqualifies the others).

Sensinode's open \textit{NanoStack 1.1} \cite{lembo2010} was superseded by the proprietary implementation of \textit{NanoStack 2.0}, thus does not satisfy the requirements we derived in Section \ref{sec:objectives}.

A number of relevant network stacks were based on \textit{TinyOS}~\cite{levis2005}.
However, since they were using \textit{TinyOS'} exotic programming language nesC, they do not match the requirements from Section \ref{sec:objectives}.
Additional, we argue that due to the high complexity of \textit{TinyOS'} system design and therefore limited number of available developers (as analyzed by P.~Levis~\cite{levis2012}), it is very unlikely that development of these stacks will keep up with the evolution of new IoT protocols.

An interesting approach towards a fully configurable network stack for embedded environments was proposed with \textit{CiAO/IP}~\cite{borchert2012}.
However, it does not match the derived requirements for similar reasons as the stacks for \textit{TiniyOS}, since it is based on an exotic C++ dialect and an exotic compiler.
Moreover, the intended granularity of configurability is too fine grained to be manageable by most application developers.

The two most prominent embedded network stacks today are \textit{uIP}~\cite{dunkels2003} and \textit{lwIP}~\cite{dunkels2001}.
Both were developed at the same time by the same author as pure IPv4 stacks.
Over time \textit{uIP} has evolved from being developed as a stand-alone network stack to being maintained as the default network stack of the Contiki operating system, supporting a full 6LoWPAN protocol stack~\cite{dunkels2004, durvey2008}.
The stack does however not support multiple network interfaces and is further based on an event loop paradigm.
This makes it hard to program for a typical programmer experienced in traditional networking applications and more difficult to implement several protocols and mechanisms from the TCP/IP suite~\cite{draft-hahm-lwig-painless-constrained-programming}.
The \textit{lwIP} stack is similar being developed over time, IPv6 support being recently added.
For use in the IoT \textit{lwIP} is missing support for 6LoWPAN.
Although both stacks can be configured to a good extent, they are missing clear documentation and interfaces for easy extensibility.
For these reasons we see both stacks failing to comply to the derived requirements.

%Both were developed at the same time by the same author as pure IPv4 stacks, but have since evolved into IPv6 stacks with support for the 6LoWPAN suite of protocols \cite{durvey2008}. \textit{uIP} was initially developed for use without an operating system targeting class 0 devices. As of today it is the default network stack for \textit{Contiki}~\cite{dunkels2004}. On the other hand, \textit{lwIP} was initially designed for slightly larger systems offering more features and flexibility. lwIP is widely used (e.g., used on top of FreeRTOS~\cite{freertos:home}). However, while both network stacks are configurable to a certain extent while providing many features relevant to the IoT, they do support for multiple interfaces and nor cleanly defined internal interfaces for extensibility.
%
%\begin{itemize}
%\item need stronger statement why lwIP and uIP do not match our requirements
%\item What about portability of existing approaches
%\end{itemize}

\section{General Architecture}
\label{sec:architecture}

The key design rule for the proposed network stack software architecture is a strict module-driven design. 
We emphasize especially on a clean definition of the interfaces between these software modules as this ensures interchangeability of modules (i.e. to choose from different implementations for different scenarios) and interoperability of these modules.
In this section we will give a brief overview on the most relevant design decisions.

\subsection{Modular Design}
\label{sec:architecture_modulardesign}

The top level of the software architecture consists of a number of high-level modules, one for each functional entity of the network stack, for example UDP, IPv6, 6LoWPAN, or RPL.
The novelty of the proposed architecture is that each high-level module is executed in its own thread while each module services the same API utilizing the operating systems IPC.
The unified interfaces allows for chaining multiple modules together and the concept is comparable to Unix \textit{STREAMS}, as proposed in the 80s \cite{ritchie1984}, with the difference that we transferred the \textit{STREAMS} concept to work via IPC.

\begin{figure}%[htb!]
  \centering
  \includegraphics[width=1\columnwidth]{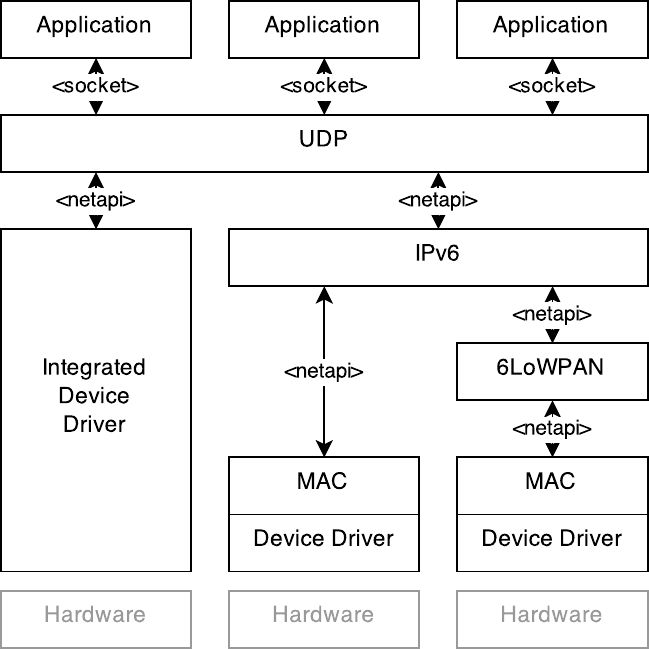}
  \caption{Sample configuration of a network stack. Each box depicts a high-level module running in it's own thread.}
  \label{fig:general_layers}
\end{figure}

Figure \ref{fig:general_layers} illustrates as network stack configuration with three devices. 
The \textit{netapi} depicts the unified IPC API between the high-level modules.
Although each of these modules can roughly be mapped to layers of the TCP/IP model, the architecture does not enforce this mapping.

This design allows for a very flexible configuration of modules (even at run-time if needed) and, as important, it enables a straight-forward extension by new features or adding other layers.
During design and implementation of modules this design enables further a clear separation of concerns and it enables for efficient testing of the modules.
Using a unified IPC API yields further benefits when adding integrated network devices into the system that include already parts of the protocol stack, like Texas Instrument's CC3000 which already provides a full TCP/IP stack~\cite{ticc3000:wiki}.
For a given network interface that e.g. already includes a full IP implementation one simply needs to write a host-side device driver that can service  \textit{netapi} and make it known to one or more transport layer modules.

One might argue that IPC comes with a high price w.r.t. run-time performance and therefore energy usage. However, our measurements using RIOT on state-of-the-art IoT hardware (a 32-bit ARM Cortex-M3 platform) show that sending a message from one thread to another, including context save, running the scheduler and context restore, requires a number of CPU cycles that is only one order of magnitude more compared to the number of cycles needed for a direct function call. The benefits thus outweigh this overhead because (i) packet throughput on IoT devices is typically low, and (ii) there are few layers going up the stack, typically yielding IPC on less than 4 occasions.

\subsection{Inter-module Communication: netapi}
\label{sec:architecture_netapi}

We introduce a unified interface for communication between high-level modules called \textit{netapi}. 
This interface is built around a small set of messages sent between the modules utilizing the operating systems IPC. The idea behind this interface is that every layer in the network stack services an identical interface. 
The core of the \textit{netapi} interface is a minimal set of messages, of which the most essential are \textit{WRITE\_DATA}, \textit{REGISTER\_RECEIVE\_CALLBACK}, and \textit{SET\_ and GET\_OPTION}.
As each module must be able to parse the general format of \textit{netapi} messages, it can implement any subset of possible message types and reply with an \textit{ENOTSUP} ("Operation not supported", \emph{POSIX.1-2001}) for all other message types~\cite{ieee:posix.1-2001}.

\subsection{Driver Interaction: netdev}
\label{sec:architecture_netdev}

The proposed architecture introduces a second unified interface for communication between device driver and medium access control (MAC) protocols, called \textit{netdev}. 
In contrary to the \textit{netapi} interface this API is based on direct function calls instead of IPC.
The practical reason for introducing a second interface at this stage are the tight timing constraints of MAC protocols (e.g. schemes based on TDMA).
Using the \textit{netdev} API allows (i) for independent implementations of device drivers and MAC protocols and (ii) for better re-use and exchangeability of both, subsequently increasing the portability.

\subsection{Packet Buffering}
\label{sec:architecture_packetbuf}

A key issue to solve in the design of a network stack for constrained devices is the handling of buffers for user data and protocol headers, as these are stored in RAM being one of the most limited resources.
Typical design choices for these buffers include centralized approaches, copying data from module to module as well as mixed concepts.
The data handling in the proposed network stack is designed around a 'copy twice' concept.
Outgoing data is copied once from the user application (socket) into a central buffer and once into a network interface's device buffer by the device driver.
The same is true for receiver data, which is copied on arrival once from the network interface into the central buffer and once more when handed over to an application.

The central packet buffer is designed as a central module accessible from all high-level modules through a well defined API.
The buffers task is to centrally provision memory for storing header and user data while it is passing through the network stack, either as packets in one piece or as fragments.
By accessing the packet buffer though a defined interface, it is further possible to transparently exchange the packet buffers implementation at compile time, e.g. one that manages a fragment of statically allocated memory against an implementation using dynamic memory on the heap.

The major advantages of a central buffer are (i) flexibility,  (ii) efficiency through less data copying and (iii) the possibility to globally define the (maximum) amount of memory used.
A drawback of a buffer taking chunks of data in different sizes is fragmentation, but we argue that with efficient implementation this disadvantage is marginal.
By including means of prioritization for memory allocations in the packet buffers API, we can further make sure that no network module is being starved by missing buffer space, thus removing the major source for dead-locks.

\section{Preliminary Evaluation}
\label{sec:evaluation}

We implemented a proof of concept of our approach for the operating system RIOT~\cite{riot:github}. To illustrate the principle feasibility we present the required amount of memory. Note that the values are still subject to optimization.

Our evaluation is based on a simple configuration using UDP, 6LoWPAN, and a single IEEE802.15.4 network interface built for the IoT-LAB\_M3 hardware~\cite{iotlab:home}. Table~\ref{tab:codesize} shows the ROM usage for relevant modules of the network stack. Our modular network stack, which is based on common programming techniques and system calls, requires less than 30~kByte of ROM and is thus in line with IoT resources.

\begin{table}[h]
\footnotesize
\begin{tabular}{ lccccc}
\toprule
\multirow{2}{*}{Module}  & IEEE & 6LoWPAN & \multirow{2}{*}{UDP}  & Socket & Helper\\
& 802.15.4 & and IPv6 & & API & Functions 
\\ \midrule
Bytes & 1,112 & 15,708 & 886 & 1,280 & 2,530
\\ \bottomrule
\end{tabular}
\caption{Preliminary code size of main network stack modules on an IoT-LAB\_M3 node (ARM Cortex-M3)}
\label{tab:codesize}
\end{table}

The RAM usage is mainly driven by two factors: (i) buffers and (ii) stacks. 
While the size for the central packet buffer is dynamically configurable during compile time, we estimate that networks like IEEE802.15.4 require less than 1-2~kByte of RAM. 
The memory consumed by stacks is dependent on the number of high-level network modules that are configured. In our setup, we use one thread per network function (i.e., UDP, IP, 6LoWPAN, and the link-layer). 
With a default stack size of 1~kByte for ARM Cortex-M3 platforms, this estimates to an additional memory usage of 4~kByte. Overall, the required RAM size complies with the target platforms (i.e., $<$ 10~kByte).

\section{Conclusion \& Outlook}
\label{sec:conclusions}

In this paper, we questioned the applicability of current network stack solutions for the Internet of Things (IoT). 
Following the observation that several IoT scenarios introduce constrained devices but do not require an ultimate memory-efficient network stack, we elaborate the design space and introduce a software architecture for a modular, full-featured network stack. 
Our proof of concept is based on a common system environment and requires $<$ 10~kBytes of RAM and $<$ 22~kBytes of ROM.
Our next steps will be to complete our implementation for the open source operating system RIOT and explore the limits of our concept in different IoT scenarios.

\newpage

{\footnotesize \bibliographystyle{acm}
\bibliography{references,rfcs}}

\end{document}